\documentclass[manuscript=article,layout=twocolumn]{achemso}

\usepackage[T1]{fontenc}

\usepackage{graphicx}
\usepackage{hyperref}
\usepackage{subcaption}
\usepackage{tabularx}
\usepackage{booktabs}
\usepackage{xcolor}
\usepackage{float}
\usepackage{siunitx}
\usepackage[normalem]{ulem} 
\usepackage{chemformula} 
\usepackage{amsmath}
\usepackage{amssymb}
\usepackage{caption}
\usepackage{cuted} 

\newfloat{scheme}{htbp}{los}
\floatname{scheme}{Scheme}
\floatname{chart}{Chart}
\newfloat{graph}{htbp}{loh}

\newcommand{\round}[2][1]{\num[round-mode=places, round-precision=#1]{#2}}
\newcolumntype{Y}{>{\centering\arraybackslash}X}
\title{Compressibility and High-Pressure Structure of CaMg$_2$Bi$_2$ and YbMg$_2$Bi$_2$}

\author{Mario Calderón-Cueva}
\altaffiliation{Current Address: Mechanical and Aerospace Engineering, The George Washington University, Washington, D.C. 20052, USA}
\affiliation{Chemical Engineering and Materials Science Department, Michigan State University, East Lansing, MI 48824, USA}

\author{Allison Pease}
\affiliation{Department of Earth and Environmental Sciences, Michigan State University, East Lansing, MI 48824, USA}

\author{Cheng Peng}
\affiliation{Chemistry Department, Michigan State University, East Lansing, MI 48824, USA}

\author{Wanyue Peng}
\affiliation{Chemical Engineering and Materials Science Department, Michigan State University, East Lansing, MI 48824, USA}

\author{Megan Rylko}
\affiliation{Chemical Engineering and Materials Science Department, Michigan State University, East Lansing, MI 48824, USA}

\author{Weiwei Xie}
\affiliation{Chemistry Department, Michigan State University, East Lansing, MI 48824, USA}

\author{Susannah M. Dorfman}
\affiliation{Department of Earth and Environmental Sciences, Michigan State University, East Lansing, MI 48824, USA}

\author{Alexandra Zevalkink}
\affiliation{Chemical Engineering and Materials Science Department, Michigan State University, East Lansing, MI 48824, USA}
\email{alexzev@msu.edu}

\begin{document}
% <--- UPDATED STRIP ENVIRONMENT --->
\begin{strip}
\begin{center}
\textbf{Abstract}
\end{center}

% \vspace{0.5em}

\begin{quote}
\noindent Compounds with the formula $AM_2X_2$ in the CaAl$_2$Si$_2$ structure type have garnered increasing interest across various solid-state research domains, such as quantum topological and thermoelectric materials.
Prior studies have identified high-pressure phase transitions in several compounds, including Mg$_3$Sb$_2$, Mg$_3$Bi$_2$, CaMn$_2$Bi$_2$, and SrAl$_2$Si$_2$.
In this study, we investigate the structural behavior of CaMg$_2$Bi$_2$ and YbMg$_2$Bi$_2$ under varying pressure conditions.
We synthesized crystals using the molten metal flux method and examined them through single-crystal synchrotron X-ray diffraction, employing diamond anvil cells to exert pressures up to 20 GPa.
Our analysis reveals insights into the anisotropic compressibility of these materials, highlighting the more compressible and flexible octahedral $A$-Bi bonds as the primary contributors to this anisotropy.
Moreover, we observed a phase transition in both CaMg$_2$Bi$_2$ and YbMg$_2$Bi$_2$ at pressures above 9.6 GPa and 8.7 GPa, respectively.
The newly identified high-pressure phase exhibits a distortion of the original CaAl$_2$Si$_2$ structure with space group $C\mathrm{2}/m$.
This high-pressure structure is distinct from that of related compounds (e.g., CaMn$_2$Bi$_2$, MgMg$_2$Bi$_2$), the latter exhibiting a square pyramidal coordination for the $M$ site.
\end{quote}
\end{strip}
% <--- END UPDATED STRIP ENVIRONMENT --->

\section{Introduction}
In recent years, compounds in the $AM_\mathrm{2}X_\mathrm{2}$ stoichiometry that crystallize in the CaAl$_2$Si$_2$ structure type have been extensively studied for their thermoelectric performance \cite{gascoin2005zintl,wang2009synthesis,wood2018observation,shuai2016higher}, quantum topological states \cite{chang2019realization,feng2022prediction}, and magnetic properties \cite{sangeetha2016strong,berry2022type}. 
Compounds that form the CaAl$_2$Si$_2$ structure type are highly tunable, as each \textit{A}, \textit{M}, and \textit{X} sites can accommodate a wide range of elements ($A$ denotes rare-earth or alkaline-earth elements, $M$ represents transition metals or group-13 elements, and $X$ signifies group-14 or group-15 elements). The inherent tunability of this structure has inspired research efforts to explore the stability limits of this structure as well as the transport properties \cite{peng2018crystal}.

For thermoelectric materials, low lattice thermal conductivity, $\kappa_L$, is a necessary characteristic. In the limit of phonon transport limited by Umklapp scattering,  $\kappa_L$ $\propto$ $v^2$ \cite{toberer2010zintl}, where $v$ is the velocity of collective lattice vibrations (\textit{phonons}). It is often convenient to approximate the phonon velocity, $v$, as the low-frequency wave propagation speed, i.e. the \textit{speed of sound}, $v_s$, which in turn, is related to the bond stiffness, $k$, and mass of the atoms, $M$ by $v_s$ $\sim$ $\sqrt{k/M}$ \cite{zeier2016thinking}. From these relations, it becomes evident that intrinsic low $\kappa_L$ arises from soft bonds (low $k$) and/or high mass $M$. A number of prior studies\cite{peng2018crystal,Ding2020DFT,singh2013electronic,may2012thermoelectric} have investigated bond stiffness in $A$Mg$_2Pn_2$ compounds via either acoustic measurements or via first principles lattice dynamics calculations, showing that the $A$-site cation plays an important role. Comparisons of compounds with \textit{A} = Mg versus \textit{A} = Ca or Yb have shown that the larger cations (Ca and Yb) lead to higher experimental elastic constants and lower anharmonicity\cite{peng2018unlikely}. However, only a few studies \cite{kundu2022topological,sun2017computational,sun2017thermoelectric} report their thermoelectric and topological properties, specifically for the CaMg$_2$Bi$_2$ and YbMg$_2$Bi$_2$ compounds. A less common approach to measuring bond stiffness is through the application of high pressure. By measuring the way in which the crystal structure evolves under applied pressure, one can simultaneously measure the compressibility of a solid and explore its stability limit with respect to high-pressure polymorphs \cite{walsh2018high,badding1998high}. High pressure also can be used to tune electronic and thermal properties of a material, making it \textcolor{black}{ a valuable tool for investigating transport phenomena \cite{zhang2023review}}. 

Although a large number of compounds form the CaAl$_2$Si$_2$ structure (space group $P\mathrm{\bar{3}}m1$), there have only been a handful of high-pressure crystallographic investigations. These initial studies hint at a surprising diversity of high pressure structure transitions: SrAl$_2$Si$_2$ \cite{zevalkink2017making, strikos2020pressure} and SrMn$_2$P$_2$  \cite{xie2017tetragonal} transform into the well-known tetragonal ($I\mathrm{4}/mmm$) crystal structure, while CaMn$_2$Bi$_2$ \cite{gui2019pressure} and MgMg$_2$(Sb,Bi)$_2$ \cite{calderon2021anisotropic} were each reported to transform into a different monoclinic structure (with space group $P\mathrm{2_1}/m$ and $C\mathrm{2}/m$, respectively).  
The focus of the present work is on the two compounds, CaMg$_2$Bi$_2$ and YbMg$_2$Bi$_2$.  Using in-situ single crystal diffraction, we show that these compounds transform into a new monoclinic structure type, closely related to the one reported for CaMn$_2$Bi$_2$. Further, we track the evolution of the crystal structure as a function of pressure to extract volumetric, linear, and bond-specific compressibility of the CaAl$_2$Si$_2$ structure.  The present work aims to build on our prior work on MgMg$_2$Bi$_2$ \cite{calderon2021anisotropic}, by exploring how replacing Mg with larger Ca and Yb cations affects the stability of the ambient pressure phase and the details of the high pressure structure.

\par
%%% NEW Fig. 1 
\begin{figure*}[!htb]
    \centering
    \includegraphics[width=16cm]{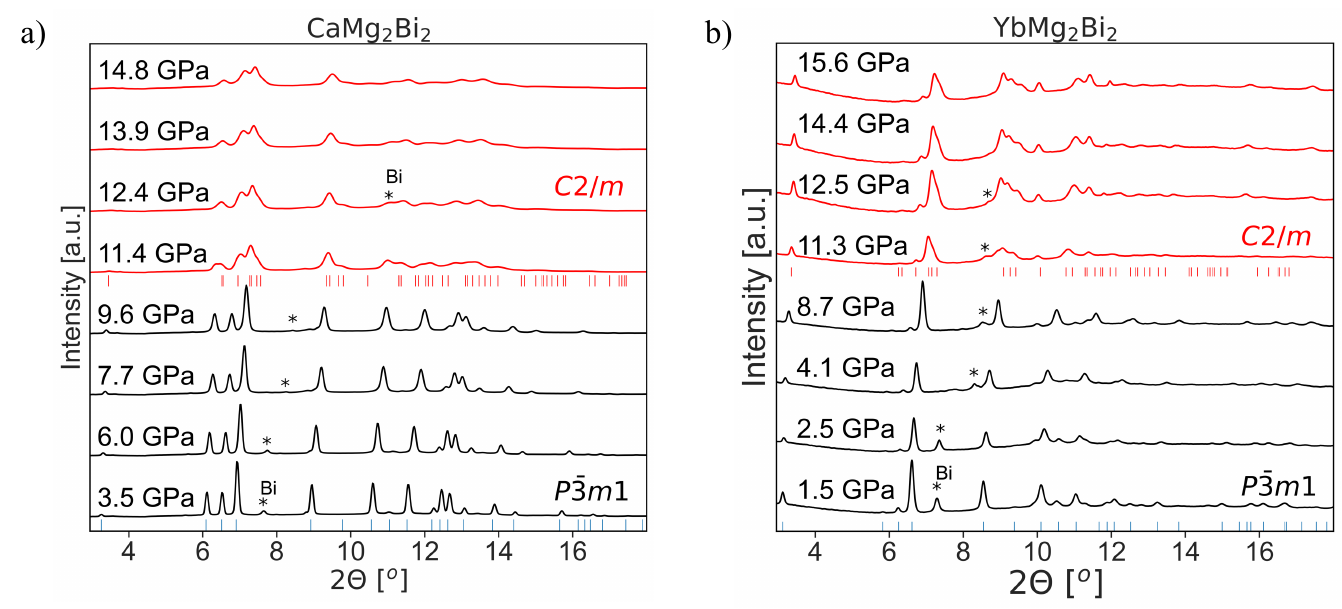}
    \caption{Powder XRD for a) CaMg$_2$Bi$_2$ and b) YbMg$_2$Bi$_2$ as a function of pressure shows evidence of phase transitions in the range of 9.6-11.4 GPa and 8.7-11.3 GPa, respectively. The trigonal phase ($P\mathrm{\bar{3}}m\mathrm{1}$) is shown in black, and the high pressure monoclinic ($C\mathrm{2}/m$) phase is shown in red.  Elemental Bi is present as an impurity in both samples, marked by asterisks.}
\label{fig:waterfall}
\end{figure*}
%%%
%%%
\section{Experimental Methods}

\subsection{Polycrystalline Synthesis}
For both the synthesis of single- and poly-crystals, the following high-purity elements were used: Ca (dendritic pieces, 99.9\%, Sigma-Aldrich), Mg (granules, Alfa Aesar 99.8\%), Sb (shot, Alfa Aesar, 99.999\%),  Bi (shot, 99.99\%, RotoMetal), and Yb (pieces, Sigma-Aldrich, 99.9\%). 
Polycrystalline samples of CaMg$_2$Bi$_2$ and YbMg$_2$Bi$_2$ were synthesized by mixing stoichiometric amounts of their corresponding elements followed by spark plasma sintering (SPS). The elements were cut into small pieces and loaded into steel vials together with stainless-steel ball bearings inside an Argon-filled glovebox. The airtight vials were then placed into a SPEX mill and ball-milled for 60 minutes.  After milling, the vials were returned to the glovebox and the fine powder was loaded into graphite dies for SPS using a Dr. Sinter SPS-211LX. 

The sintering procedure for both compounds was the same, where the powders were subjected to a pressure of 31 MPa, heated to 700$\mathrm{^{\circ}}$ C at a rate of 68 $\mathrm{^{\circ}}$C/min, and held at that temperature for 10 minutes. The target SPS temperature was selected to be 80\% of the melting temperatures reported in Ref. \cite{may2012thermoelectric}, while the pressure and wait time were empirically determined to yield the desired phase. After that period, the pressure was immediately released and the samples were allowed to reach ambient temperature. The resulting consolidated disks were ground to fine powders and checked for phase purity using X-ray diffraction (XRD) with a Rigaku SmartLab Diffractometer with a Cu $\mathrm{K\alpha}$ source. The results of the Rietveld refinements, depicted in the SI Figure \ref{SI_fig:Rieveld_AP}, revealed that the impurity content (elemental Bi) was less than 3\% and 5\% for CaMg$_2$Bi$_2$ and YbMg$_2$Bi$_2$, respectively. \par
%%%
\subsection{Single-Crystal Synthesis}
Single crystals of CaMg$_2$Bi$_2$ and YbMg$_2$Bi$_2$ were grown via the self-flux method, with a starting composition of \textit{A}Mg$_4$Bi$_6$ (\textit{A}=Ca, Yb). \textcolor{black}{The high quality of the crystals was confirmed by sharp diffraction spots in the precession images, as demonstrated in Figure \ref{SI_fig:precession} (Supplemental Information).}
After the ball-milling process, which was identical to the polycrystalline samples, the powders were then loaded into Canfield Al$_2$O\textsubscript{3} crucibles and subsequently sealed in quartz ampules under vacuum ($\mathrm{<10^{-4}}$ Torr). The quartz ampules were then placed in a box furnace, where the furnace temperature was brought to 900$\mathrm{^{\circ}}$ C in 8 hr, then to 850$\mathrm{^{\circ}}$ C in 1 hr, and cooled to 650$\mathrm{^{\circ}}$ C in 40 hours. After soaking for 48 hours at 650$\mathrm{^{\circ}}$ C, the ampules were centrifuged to remove the liquid flux from the desired crystallized samples. As a result, large crystals (approx. 5 mm in their longest dimension) were obtained. These crystals were then easily cleaved into smaller pieces with surgical blades, which yielded smaller pieces with plate-like geometry. The pieces that exhibited faceted morphology, and approx. 50 $\mathrm{\mu m}$ in their longest dimension, were selected and screened for crystallinity and phase purity using a Rigaku XtaLAB Synergy S Diffractometer. \par
%%%
\subsection{High-pressure X-ray diffraction}
High-pressure diffraction experiments were conducted at the Advanced Photon Source (APS) in Argonne National Lab, using diamond anvil cells (DACs) to apply hydrostatic pressure environments to the samples. \textcolor{black}{X-ray diffraction patterns were collected during the compression cycle}.\par
For powder diffraction of YbMg$_2$Bi$_2$, pairs of flat, 800-$\mathrm{\mu m }$-culet diamond anvils (BX90 \cite{kantor2012bx90} design) were used. A precompressed powder of the desired composition was loaded into a 410-$\mathrm{\mu m}$ hole within a pre-indented steel gasket. A methanol-ethanol (4:1) mixture served as a pressure transmitting medium, and a 10-$\mathrm{\mu m}$ ruby crystal was used as the pressure indicator. The diffraction patterns were collected at beamline 13-BM-C, \textcolor{black}{using X-rays with a wavelength $\lambda$ =  0.413 Å}. CaMg$_2$Bi$_2$ powder diffraction experiments were carried out at APS beamline 16 BM-D, \textcolor{black}{ with X-ray light of $\lambda$ =  0.434 Å}, using flat, 300-$\mathrm{\mu m }$-culet diamond anvils (BX90 design), and Neon as the pressure transmitting medium, using the COMPRES-GSECARS gas loading system\cite{Rivers2008}. A pre-indented rhenium gasket was used, in which a hole of $\mathrm{\sim}$ 150 $\mu m$ in diameter was drilled. 
\par
\begin{figure*}[!htb]
    \centering
    \includegraphics[width=14cm]{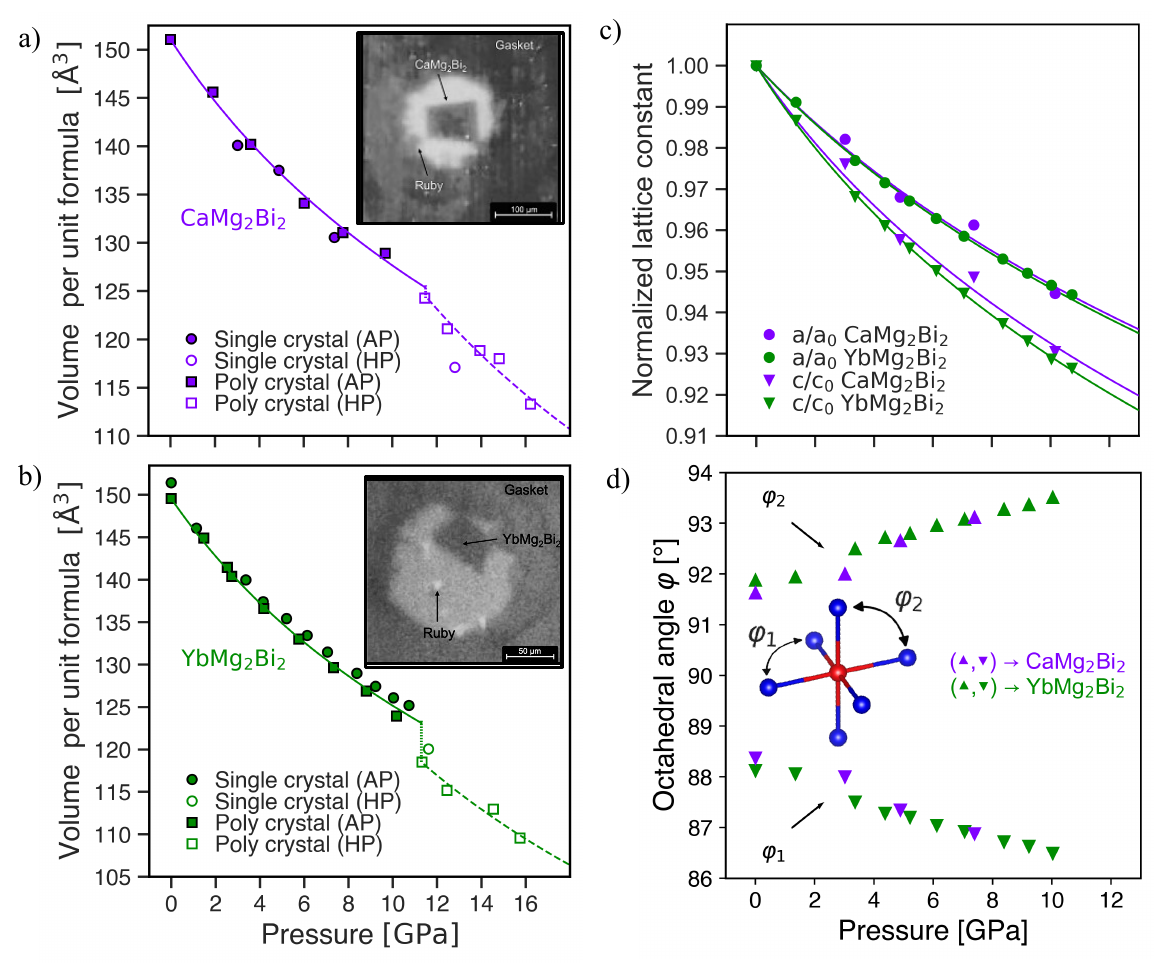}
    \caption{\textcolor{black}{Experimental single-crystal (circles) and polycrystalline (squares) unit cell volumes as a function of pressure for a) CaMg$_2$Bi$_2$ and b) YbMg$_2$Bi$_2$ and their corresponding images inside DACs \textcolor{black}{after pressure-medium-loading (see insets, methanol-ethanol for CaMg$_2$Bi$_2$ and Ne for  YbMg$_2$Bi$_2$)}. The open symbols in a) and b) show the volume/formula unit of the high-pressure phase.  c) The normalized lattice parameters d) and octahedral angles vs. pressure for CaMg$_2$Bi$_2$ (magenta) and YbMg$_2$Bi$_2$ (green) from SC-XRD.  Dashed lines represent 2\textsuperscript{nd} order BM fits. 
    }
    }
    \label{fig:compress}
\end{figure*}

For single crystal experiments, diamond anvil pairs of flat 300-$\mathrm{\mu m}$ culets (custom CVD, conical Boehler-Almax design, with (Vascomax\textsuperscript{\textregistered}) stainless steel were used,  with an effective aperture of 80$\mathrm{^\circ}$. Pre-screened single crystals of the desired compositions were loaded into a 150-$\mathrm{\mu m}$ hole within a pre-indented Re gasket. Ne gas or a methanol-ethanol (4:1) mixture was used as the pressure transmitting medium, and a 10-$\mathrm{\mu m}$ ruby crystal was used as the pressure indicator. Ne gas was loaded at GSECARS (Sector 13) \cite{Rivers2008} at APS. Single-crystal X-ray diffraction (SC-XRD) experiments were carried out in beamline 13-BM-C, \textcolor{black}{with X-ray wavelength source of $\lambda$ =  0.434 Å}. The inset images in Figure \ref{fig:compress}a-b) show the loading conditions of the CaMg$_2$Bi$_2$ and YbMg$_2$Bi$_2$ crystals inside the DACs after gas-loading.
The diffraction patterns were collected with a Dectris Pilatus3 1M Pixel Array Detector. \par
%%%
The diffraction data for single-crystalline CaMg$_2$Bi$_2$ and YbMg$_2$Bi$_2$ was collected by APEX3 \cite{bruker2016apex3}, integrated using SAINT \cite{bruker2009saint} code and corrected for absorption with the SADABS method \cite{krause2015comparison}. The structure was solved by the SHELXS routine \cite{sheldrick2008short}. For both compositions, the structures were refined in the OLEX2 Suite \cite{dolomanov2003olex}, by the full-matrix least-square methods of SHELXL. \textcolor{black}{All refined structural data was obtained from the single crystal samples. The powder samples were used only for initial determination of the phase transition pressure, and Rietveld fits to the high-pressure phases (see SI Figure \ref{SI_fig:Rieveld_HP}) were used as an additional verification of the structural model obtained from single crystal refinements}. Rieveld refinements were performed in GSAS-II \cite{toby2013gsas}.\par
%%%
After the crystallographic refinements, the parameters of the equations of state (EoS) for the two compounds were fit to data via the least-squares fitting routine of the EoSFit7 software\cite{angel2014eosfit7c}. The volume compressibilities were fit using a second-order Birch-Murnagham (BM) EoS \cite{birch1947finite, murnaghan1944compressibility}. Polyhedral volumes and uncertainties were calculated from bond lengths using the approach described in the Supplemental Information section \ref{SI_sec:error}.\par
%%%

\section{Results and Discussion}

\textcolor{black}{The powder and single crystals of CaMg$_2$Bi$_2$ and YbMg$_2$Bi$_2$ form the trigonal CaAl$_2$Si$_2$ structure (space group $P\mathrm{\bar{3}}m1$, as shown in Figure \ref{fig:P-3m1}b), as reported in prior literature \cite{deller1977ternare,may2011structure}}.
The powder XRD patterns on polycrystalline samples were initially collected for CaMg$_2$Bi$_2$ and YbMg$_2$Bi$_2$  as a function of pressure up to 18 GPa and 17.5 GPa, respectively.   Figure  \ref{fig:waterfall} shows the patterns at selected pressures. In both samples, the trigonal phase is shown in black. New peaks begin to appear in the diffraction patterns of CaMg$_2$Bi$_2$ and YbMg$_2$Bi$_2$ above 9.6 and 8.7 GPa, respectively, indicating the beginning of a phase transition. \par

 Rietveld analysis of the powder diffraction data above 11.3 GPa was initially attempted using two known high-pressure structures of similar compounds, namely that of $C\mathrm{2}/m$-Mg\textsubscript{3}Sb$_2$ \cite{calderon2021anisotropic} and $P\mathrm{2}_1/m$-CaMn$_2$Bi$_2$ \cite{gui2019pressure}.  However, these solutions did not provide an acceptable fit. For that reason, single-crystal XRD experiments were conducted for both compositions, ultimately showing that CaMg$_2$Bi$_2$ and YbMg$_2$Bi$_2$ crystallize in a new structure type with space group $C\mathrm{2}/m$. In the following sections, we present first an analysis of the compressibility of each compound within the stability range of the trigonal phase, followed by a description of the high-pressure structures of both compounds, based on the single-crystal data. Tables \ref{tab:crystallography_CaMg2Bi2}-\ref{tab:crystallography_HP} in the Supplemental Information summarize the relevant experimental data for the Birch-Murnagham fits to single-crystal data obtained for both compounds. These tables show selected pressure steps for both the ambient- and high-pressure phases.

\begin{figure}[!htb]
    \centering
    \includegraphics[width=0.45\textwidth]{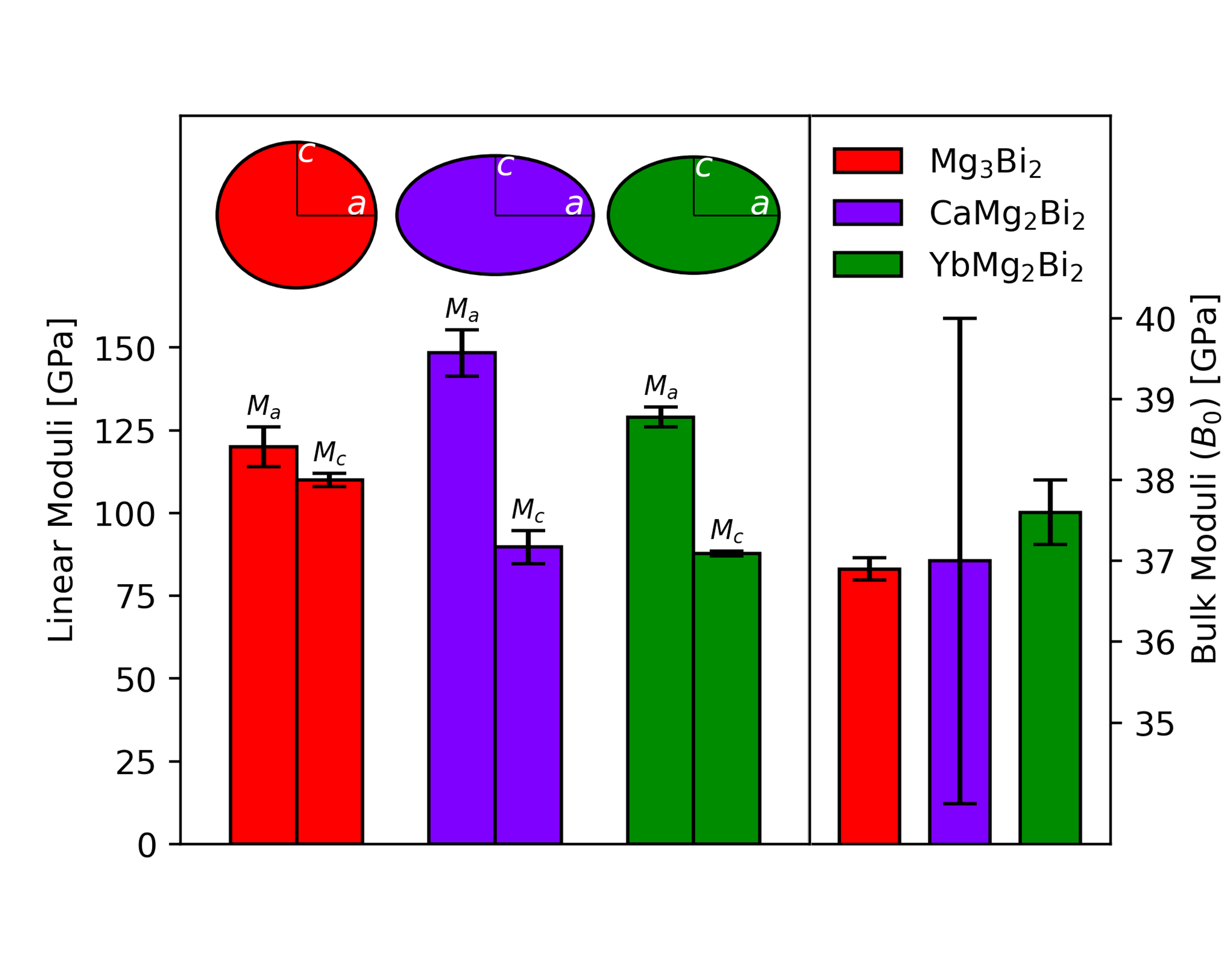}
    \caption{Linear (left) and bulk (right) moduli comparison for $A$Mg$_2$Bi$_2$ ($A$ = Ca, Mg, Yb) at ambient pressure with their corresponding uncertainties. MgMg$_2$Bi$_2$ data taken from Ref.  \cite{calderon2021anisotropic}.}
    \label{fig:moduli_comp}
\end{figure}

\subsection{Compressibility of CaMg$_2$Bi$_2$ and YbMg$_2$Bi$_2$}

\textit{Volume compressibility}: The experimental unit cell volume as a function of pressure is shown in Figure \ref{fig:compress}a-b) for CaMg$_2$Bi$_2$ and YbMg$_2$Bi$_2$.  Filled symbols correspond to the ambient-pressure trigonal structure type and open symbols to the high-pressure structure type, which will be discussed further in the next section.  The volumetric bulk modulus for the trigonal ambient-pressure ($B_{0}$) phases were calculated by fitting the pressure-volume data (using \textcolor{black}{both} single and polycrystalline data points) to the second-order Birch-Murnaghan (BM) equation of state \cite{poirier1998logarithmic,birch1947finite,murnaghan1944compressibility}: 
\begin{equation}
P(V) = \frac{3B_0}{2}\left[\left(\frac{V_0}{V}\right)^{\frac{7}{3}}-\left(\frac{V_0}{V}\right)^{\frac{5}{3}}\right]
\label{eq:BM_2nd}
\end{equation}
where $P$ is the applied pressure (determined from the ruby standard), $V$ the observed unit cell volume, and $V_0$ the reference unit cell volume. Note that $B_0$ corresponds to the bulk modulus at ambient pressure, and is treated here as a pressure-independent fitting parameter. The BM fits to the volume data are shown as the purple and green curves in Figure \ref{fig:compress}, and the resulting $B_0$ values are listed in Table \ref{tab:moduli}.  

CaMg$_2$Bi$_2$ and YbMg$_2$Bi$_2$ were found to have nearly identical zero-pressure $B_0$ ($\sim$ 37-38 GPa).  This is not surprising, given the similar ionic radii of Yb$^{2+}$ (1.02 \AA\cite{shannon1976revised}) and Ca$^{2+}$ (1.00 \AA\cite{shannon1976revised}), and similar ambient pressure volume.  Our results are also in agreement with previously reported values obtained using ultrasonic measurements on bulk polycrystalline samples (36 GPa and 38 GPa for CaMg$_2$Bi$_2$ and YbMg$_2$Bi$_2$, respectively \cite{peng2018unlikely}).  

At the phase transition from the trigonal to monoclinic cell, a  \textcolor{black}{small decrease in the volume per formula unit of $\mathrm{\sim}$ 0.9\% and $\mathrm{\sim}$ 3.7 \% for CaMg$_2$Bi$_2$ and YbMg$_2$Bi$_2$, respectively, can be observed}.  Above the phase transition, we have insufficient data points to reliably determine the elastic moduli of the high-pressure structure type \textcolor{black}{(the lattice parameters vs. pressure for the $C2/m$ are shown in the Fig. \ref{SI_fig:HP_lat_params})}. However, BM fits are shown as dashed curves to provide a "guide to the eye", and these were to estimate the change in volume at the phase transition pressure.

%%% New Fig. 3
\begin{figure*}[!htb]
    \centering
    \includegraphics[width=12cm]{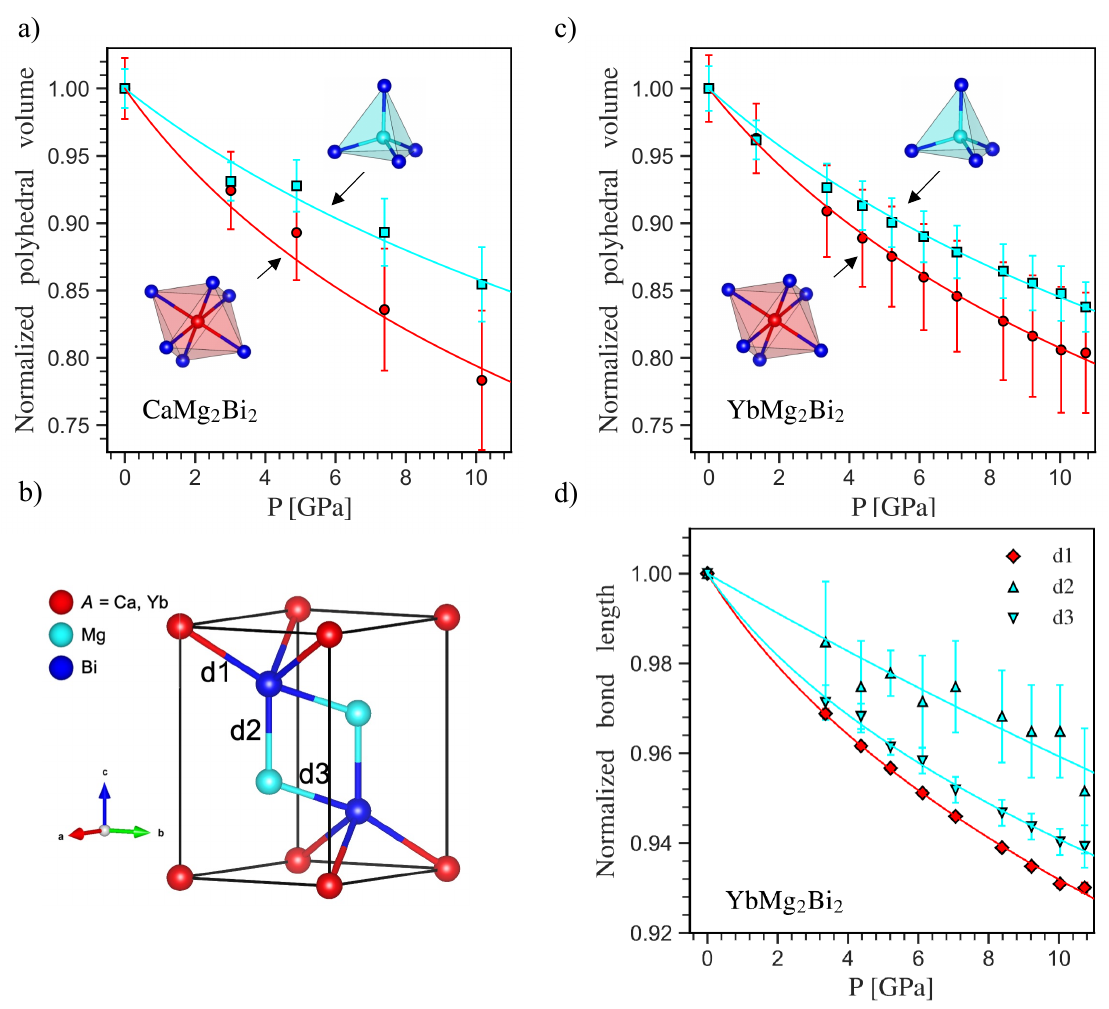}
    \caption{Normalized polyhedral volume vs. pressure below the phase transition for a) CaMg$_2$Bi$_2$ and c) YbMg$_2$Bi$_2$ from SC-XRD. b) The ambient pressure trigonal structure of \textit{A}Mg$_2$Bi$_2$ (\textit{A} = Ca, Yb), space group $P\mathrm{\bar{3}}m\mathrm{1}$, showing the octahedral ($d1$) and tetrahedral bonds ($d2$ and $d3$) d) Normalized bond length vs. pressure for YbMg$_2$Bi$_2$ from SC-XRD highlights more compressible octahedral bond.}
    \label{fig:P-3m1}
\end{figure*}
%%%

\textit{Linear compressibility}: Returning to the trigonal structure type,  Figure \ref{fig:compress}c) shows the normalized lattice parameters, $a/a_0$ and $c/c_0$, as a function of pressure, obtained from single crystal data. From these, the linear bulk moduli ($M_a$ and $M_c$) were extracted using a linearized 2$^{nd}$ order BM equation, given by: 
\begin{equation}
    P(l) = \frac{3M_l}{2}\left[\left(\frac{l_0^3}{l^3}\right)^{\frac{7}{3}}-\left(\frac{l_0^3}{l^3}\right)^{\frac{5}{3}}\right].
    \label{eq:BM_2nd_linear}
\end{equation}
\textcolor{black}{Here, $l$ represents either the lattice parameter $a$ or $c$, \cite{angel2000equations}}. The values at ambient pressure, $a_0$ and $c_0$, were treated as fitting parameters.  Figure \ref{fig:compress}c) clearly shows that $c$ is more compressible than $a$ for both compounds.  Accordingly, the linear moduli obtained from the BM fits along $c$ are much smaller ($M_c$ $\sim$ 90 GPa) than those obtained along $a$ ($M_a$ $\sim$ 130-140 GPa). Figure \ref{fig:moduli_comp} shows a comparison of the bulk and linear elastic moduli of MgMg$_2$Bi$_2$, CaMg$_2$Bi$_2$, and YbMg$_2$Bi$_2$. The ellipses illustrate how the linear modulus varies as a function of crystallographic direction in the a-c plane.  Conceptually, they can be understood as the final cross sections of spherical particles subjected to isostatic pressure.  We can see that even though the bulk moduli are nearly identical regardless of cation species, the degree of anisotropy varies strongly. Compared with MgMg$_2$Bi$_2$, the larger cations in the present compounds lead to decreased stiffness (smaller $M_c$) along the c-axis, and increased stiffness along the a- and b-axes (larger $M_a$). This would be expected to lead, in turn, to anisotropy in the thermal transport in single crystals.   

\begin{table*}[!htb]
\caption{Summary of the fitted ambient-pressure moduli for the trigonal CaMg$_2$Bi$_2$ and  YbMg$_2$Bi$_2$ compounds. Bulk modulus ($B_0$), linear moduli in $a$- and $c$-directions ($M$), and polyhedral moduli ($K_{tet}$, $K_{oct}$), and bond-specific moduli ($K_{d1}$, $K_{d2}$, $K_{d3}$) are included.} %Ambient-pressure values of $V$, $a$, $c$, $V_{tet}$, $V_{oct}$, $d1$, $d2$, and $d3$ are obtained from the best fit of all data.}
\centering
\setlength\tabcolsep{1.5pt}
\begin{tabularx}{\textwidth}{*{3}{Y}}
                                  & CaMg$_2$Bi$_2$                    & YbMg$_2$Bi$_2$                     \\
\toprule
Zero-pressure moduli  (GPa)          &                              &                              \\ \midrule
Bulk modulus (AP), $B_{0}$              & 37 $\mathrm{\pm}$ 3      & 37.6 $\mathrm{\pm}$ 0.4  \\
Linear modulus along a, $M_a$ & 148.4 $\mathrm{\pm}$ 7   & 129.0 $\mathrm{\pm}$ 3   \\
Linear modulus along c, $M_c$ & 89.7 $\mathrm{\pm}$ 5    & 87.8 $\mathrm{\pm}$ 0.8  \\
Tetrahedron modulus, $K_{tet}$  & 48 $\mathrm{\pm}$ 1      & 42.4 $\mathrm{\pm}$ 0.25 \\
Octahedron modulus, $K_{oct}$  & 27.3 $\mathrm{\pm}$ 0.15 & 30.4 $\mathrm{\pm}$ 0.1 \\ 
Modulus of bond $d1$, $K_{d1}$      & -- &  96 $\mathrm{\pm}$ 1.5  \\
Modulus of bond $d2$, $K_{d2}$      & -- &  168 $\mathrm{\pm}$ 15  \\
Modulus of bond $d3$, $K_{d3}$      & -- &  119 $\mathrm{\pm}$ 2.5 \\     
\bottomrule
\label{tab:moduli}
\end{tabularx}
\end{table*}
%%%

\textit{Compressibility of polyhedra and individual bonds:} To better understand the underlying factors controlling the anisotropic compressibility, we consider how individual bond lengths and bond angles evolve as a function of pressure \cite{hazen1979bulk}. As shown in Fig. \ref{fig:P-3m1}b), the trigonal \textit{A}Mg$_2$Bi$_2$ (\textit{A} = Ca, Bi) structure consists of layers of [Mg$_2$Bi$_2$]$^{2-}$, in which the Mg atom (cyan) is tetrahedrally-coordinated by Bi (blue). Each tetrahedron shares three edges with neighboring tetrahedra. The tetrahedal layers are separated by the divalent cations (red), i.e., trigonal layers of cations (Ca$^{2+}$ or Yb$^{2+}$) that are octahedrally coordinated by six Bi atoms.  The regular octahedra possess one unique bond ($d1$) only. In contrast, the tetrahedra have two distinct bonds, one apical bond ($d2$) and three basal bonds ($d3$).  Accurate Mg positions, which are needed to obtain $d2$ and $d3$ as a function of pressure, could only be refined from the YbMg$_2$Bi$_2$ SC-XRD data, but not for CaMg$_2$Bi$_2$.  Therefore, the pressure-dependent bond lengths are only shown for YbMg$_2$Bi$_2$ (Figure \ref{fig:P-3m1}d). However, the overall volume of each polyhedral environment (octahedral and tetrahedral) as well as the Bi-$A$-Bi bond angles could be obtained using only the refined Bi position, and was therefore calculated for both compounds. 

The polyhedral volumes as a function of pressure are shown in Fig. \ref{fig:P-3m1}a) \& c). The uncertainty was estimated using an error propagation method discussed in detail in the Supporting Information Section \ref{SI_sec:error}. The experimental data clearly demonstrate that, for both compounds, the octahedral environment compresses at a faster rate than the tetrahedral environment.  This can be quantified  using a modified version of Eq. \ref{eq:BM_2nd}, in which the unit cell volume is replaced by the polyhedral volume, yielding zero-pressure “polyhedra bulk moduli” (defined here as $K_{oct}$ and $K_{tet}$, listed in Table \ref{tab:moduli}) \cite{cohen1985calculation}.  The BM fits are shown as dashed curves in Fig. \ref{fig:P-3m1}a) \& c).  For both CaMg$_2$Bi$_2$ and YbMg$_2$Bi$_2$, the tetrahedron bulk modulus, $K_{tet}$, is considerably larger than the octahedron bulk $K_{oct}$.  As a result, in both compounds, the overall bulk modulus lies between that of the tetrahedral modulus and the octahedral modulus. 

Diving deeper, we find that the octahedral environment is more compressible than the tetrahedral one for two reasons; first, the octahedral bonds are more compliant, and second, the octahedral angles vary with pressure, allowing for further volume compression.  The former can be seen in Figure \ref{fig:P-3m1}d), which shows $d1$, $d2$, and $d3$ as a function of pressure in YbMg$_2$Bi$_2$.  Note that uncertainty in the positions of the light Mg atoms leads to large uncertainty in $d2$ and $d3$.  Initial values of bond lengths for YbMg$_2$Bi$_2$ are $d1$ = 3.2838(3) \AA,  $d2$ = 2.8990(3) \AA, $d3$ = 2.9689(2) \AA; %and $d1$ = 3.290(4) \AA, $d2$ = 2.92(10) \AA, and $d3$ = 2.91(4) \AA \ for CaMg$_2$Bi$_2$. 
Figure \ref{fig:P-3m1}d) clearly shows a faster compression with pressure of the octahedral bond $d1$ with respect to the other bonds.  The individual bond moduli ($K_{di}$, where i = 1, 2, or 3) were obtained using a modified version of Eq. \ref{eq:BM_2nd_linear}, and are reported in Table \ref{tab:moduli}. The $\mathrm{2^{nd}}$-order BM fits are shown as dashed lines in Fig. \ref{fig:P-3m1}d).  The octahedral bond angles, $\phi_1$ and $\phi_2$, are also allowed to vary in the trigonal symmetry. Figure \ref{fig:compress}d) shows the evolution of these two angles as a function of pressure in CaMg$_2$Bi$_2$ and YbMg$_2$Bi$_2$.  In an ideal octahedron, these angles would be 90$\mathrm{^\circ}$. At ambient pressure, the angles are 88$\mathrm{^\circ}$ and 92$\mathrm{^\circ}$ (they are nearly identical regardless of cation species). As pressure increases, these angles deviate further from 90$\mathrm{^\circ}$, indicating a distortion of the ideal octahedral geometry, allowing the structure to further compress in the $c$-direction.  This flexibiltiy, together with the more compliant octahedral bonds, explain why the $c$-axis is more compressible (softer) than the $a$-axis in this structure type.

\begin{figure*}[!htb]
    \centering
    \includegraphics[width=\textwidth]{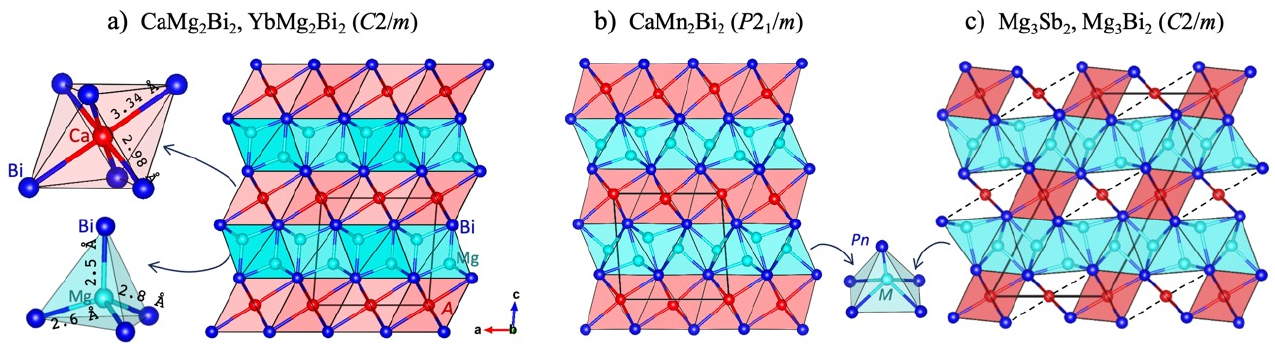}
    \caption{a) High-pressure ($C\mathrm{2}/m$) structures for CaMg$_2$Bi$_2$ and YbMg$_2$Bi$_2$ with detailed octahedron (top left) and detailed tetrahedron (bottom left). Bond lengths are given for CaMg$_2$Bi$_2$ at 12.82 GPa. In the high-pressure structures of b) CaMn$_2$Bi$_2$ space group $P\mathrm{2_1}/m$ and c) MgMg$_2Pn_2$ (\textit{Pn} = Sb, Bi), space group \textit{$C\mathrm{2}/m$}, the ambient-pressure tetrahedral environmental becomes square pyramidal. \cite{gui2019pressure} \cite{calderon2021anisotropic}}
    \label{fig:structure_comparison}
\end{figure*}

%%%%%%%%%%%%%%%%%%%%%%%%

\subsection{High-Pressure Crystal Structure}
As previously mentioned, a high-pressure structural transformation for both compounds was first observed in the powder samples at pressures between 9.5-11.4 GPa and 8.7-11.3 GPa in CaMg$_2$Bi$_2$ and YbMg$_2$Bi$_2$, respectively. \textcolor{black}{The powder data, which is more likely than single-crystal data to exhibit two-phase equilibrium, displays peak broadening that is too severe to allow for the refinement of relative phase fractions. This is relevant to CaMg$_2$Bi$_2$, where at 11.3 GPa, the ambient and high-pressure phases may co-exist.} The single crystal data confirmed these phase transition pressures. The high-pressure structures were refined as having the monoclinic crystal system with a unit cell corresponding to twice that of the $P\mathrm{\bar{3}}m1$ unit cell. The final refined space group resulting in a solution with physical interatomic distances was $C\mathrm{2}/m$. The position determination of Mg atoms for both compounds proved to be challenging due to its low mass compared to the other atoms. Several possible solutions were explored, and ultimately, the only physical solution put the Mg atoms in a coordination environment very similar to that of the ambient-pressure structure.  This new structure can be described as a slightly distorted variant of the $P\mathrm{\bar{3}}m1$ structure, in which the coordination environments are the same (i.e., alternating $A$-centered octahedral layers and Mg-centered tetrahedral layers are maintained), but the unit cell is distorted with a $\beta = 92\mathrm{^\circ}$ and $\beta = 94 \mathrm{^\circ}$ for CaMg$_2$Bi$_2$ and YbMg$_2$Bi$_2$, respectively.  The crystallographic information is summarized in Table \ref{tab:crystallography_HP}.

As shown in Fig. \ref{fig:structure_comparison}b-c), other $AM_2X_2$ compounds at high pressure, namely $\mathrm{MgMg_2}Pn_\mathrm{2}$ (\textit{Pn} = Sb, Bi) and CaMn$_2$Bi$_2$, also experience displacive phase transformations to monoclinic symmetry, but with more dramatic changes to the coordination environments than in the present compounds. In the case of MgMg$_2$Bi$_2$ and MgMg$_2$Sb$_2$, a monoclinic symmetry (Z = 4, $C\mathrm{2}/m$) was reported above 7.8 and 4.0 GPa, respectively \cite{calderon2021anisotropic}. This trigonal-to-monoclinic transformation is accompanied by octahedral bond-breaking at every other cation site, as depicted in Fig. \ref{fig:structure_comparison}c) by the dashed lines.  
In CaMn$_2$Bi$_2$, the monoclinic phase forming above 2 GPa (Z = 2, $P\mathrm{2_1}/m$) \cite{gui2019pressure}, shows instead only a slight elongation of the octahedral bonds.  However, in both of these alternative high-pressure structures (and in contrast to the subject of this study), the tetrahedral environment in the $M_2X_2$ layer is transformed into a square pyramidal one (shown by the cyan polyhedra).

\section{Conclusion}
The higher compressibility of CaMg$_2$Bi$_2$ and YbMg$_2$Bi$_2$ in the c-direction arises from the soft tetrahedral bonds, while the $a$-direction compressibility is smaller due to stiffer octahedral bonds. The $AM_\mathrm{2}X_\mathrm{2}$ compounds studied here have 3D, edge-sharing octahedra as well as tetrahedra. For this reason, to understand the  bulk modulus, it is necessary to study it as a combined effect of different bond strengths and edge-sharing 3D polyhedra. \textcolor{black}{Additionally, a high-pressure structure (monoclinic, $C\mathrm{2}/m$) was discovered above 9.6 GPa and 8.7 GPa for CaMg$_2$Bi$_2$ and YbMg$_2$Bi$_2$, respectively}. This displacive phase transition is characterized by a change in the Mg-Bi coordination environment from tetrahedral to square pyramidal coordination. The study of the elastic properties and phase stability of these compounds serves as a basis for understanding the interplay between the bonding nature and thermal transport of this class of thermoelectric materials.%

\section{Acknowledgements}
The research was supported by the U.S. Department of Energy, Office of Basic Energy Sciences, Division of Materials Sciences
and Engineering under Award DE-SC0019252. Use of the COMPRES-GSECARS gas loading system was supported by COMPRES under NSF Cooperative Agreement EAR-1661511 and by GSECARS through NSF grant EAR-1634415. This research used resources of the Advanced Photon Source, a U.S. Department of Energy (DOE) Office of Science User Facility operated for the DOE Office of Science by Argonne National Laboratory under Contract No. DE-AC02-06CH11357. \par
\textcolor{black}{The authors gratefully acknowledge Dongzhou Zhang and Jingui Xu for their help with high-pressure X-ray diffraction experiments, Sergey N. Tkachev for his help with gas loading of diamond anvil cells at APS, and Richard Staples for his support at the Center for Crystallography at Michigan State University.}
\onecolumn
\bibliography{references} 
% --- Supplementary Information Appendage ---
\clearpage
\section*{Supplementary Information}

% Reset counters for the SI section
\setcounter{equation}{0}
\setcounter{figure}{0}
\setcounter{table}{0}
\setcounter{subsection}{0}
\setcounter{page}{1}

% Prefix definitions for SI section
\renewcommand{\theequation}{S\arabic{equation}}
\renewcommand{\thefigure}{S\arabic{figure}}
\renewcommand{\thetable}{S\arabic{table}}
\renewcommand{\thepage}{S\arabic{page}} 
\renewcommand{\thesubsection}{S\arabic{subsection}}

% Hyperref anchor definitions to fix duplicate identifiers
\renewcommand{\theHequation}{S\arabic{equation}}
\renewcommand{\theHfigure}{S\arabic{figure}}
\renewcommand{\theHtable}{S\arabic{table}}
\renewcommand{\theHsubsection}{S\arabic{subsection}} 

% ---> THE FIX FOR ACHEMSO/JACS <---
% Since the JACS journal format completely disables section numbering, 
% we must manually redefine \subsection here to force it to step the 
% counter, print the "S1", "S2" label, and make \ref{} work.
\makeatletter
\renewcommand{\subsection}[1]{%
  \par\vspace{2ex \@plus 1ex \@minus .2ex}%
  \phantomsection
  \refstepcounter{subsection}%
  \noindent\textbf{\thesubsection\quad #1}\par\nobreak
  \vspace{1ex \@plus .2ex}%
}
\makeatother

%%%
\subsection{Selected crystallographic data for the ambient- and high-pressure phases}
\begin{figure}[!htbp]
    \centering
    \includegraphics[width=0.8\textwidth]{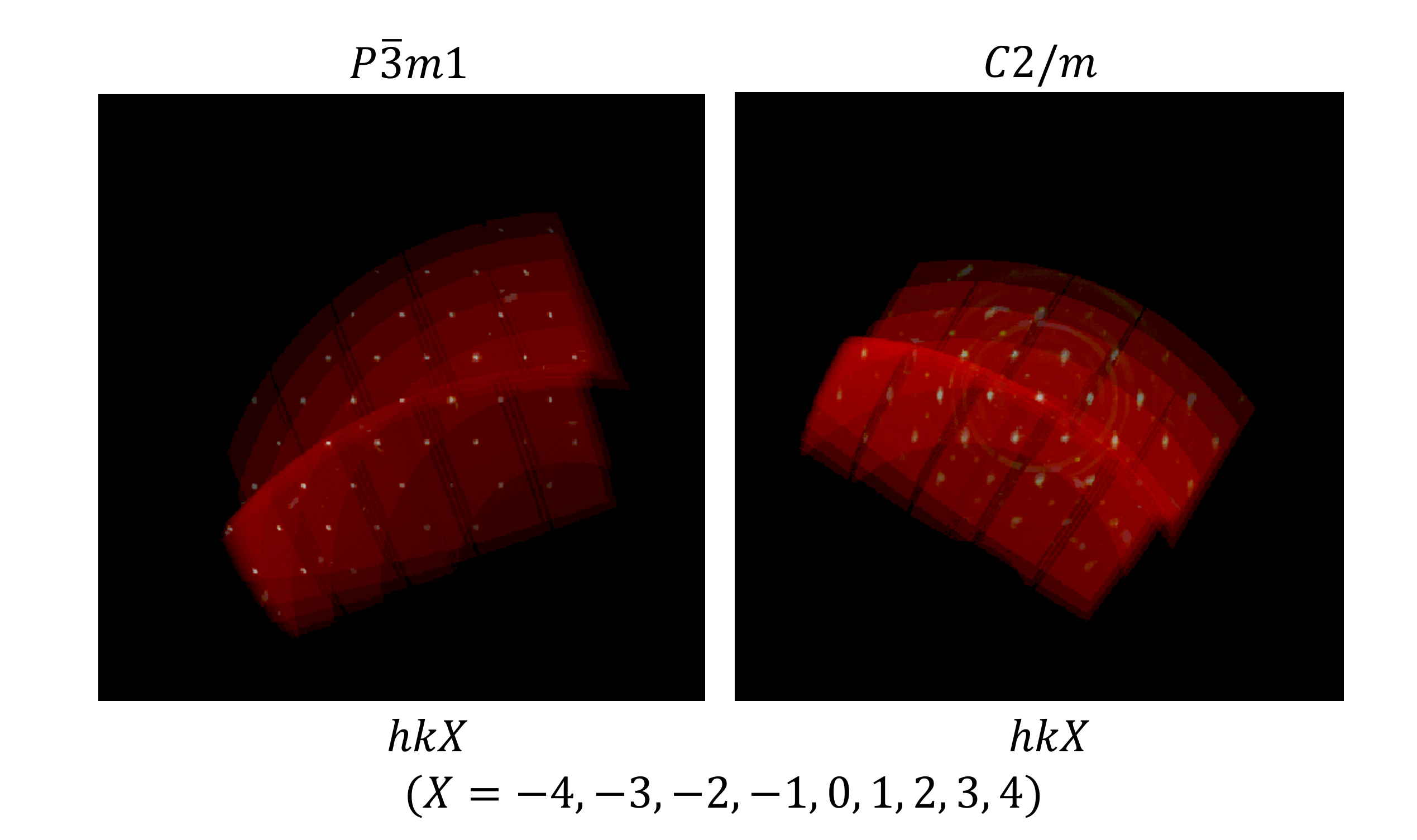}
    \caption{\textcolor{black}{Precession images for YbMg$_2$Bi$_2$ at pressures below (left) and above (right) the transition. The crystal quality below the transition pressure is excellent, while significant broadening of reflections is observed above the transition pressure. However, there is clear evidence of decreased symmetry, consistent with the powder diffraction patterns.}}
    \label{SI_fig:precession}
\end{figure}

\begin{table}[!htbp]
\caption{Crystallographic data for the trigonal CaMg$_2$Bi$_2$ structure at different pressures from single crystal XRD experiments.}
\label{tab:crystallography_CaMg2Bi2} 
\centering
\setlength\tabcolsep{1.5pt}
\begin{tabularx}{\textwidth}{l*{5}{Y}}
\toprule
Pressure (GPa)                & AP           & \round{3.02}         & \round{4.89}         & \round{7.39}         & \round{10.16}        \\ \midrule
Space Group; Z                & $P\mathrm{\bar{3}}m\mathrm{1}$; 1     & $P\mathrm{\bar{3}}m\mathrm{1}$; 1     & $P\mathrm{\bar{3}}m\mathrm{1}$; 1     & $P\mathrm{\bar{3}}m\mathrm{1}$; 1     & $P\mathrm{\bar{3}}m\mathrm{1}$; 1\\
a (Å)                         & 4.71886(17)  & 4.6483(19)   & 4.6223(15)   & 4.549(2)     & 4.472(2)    \\
b (Å)                         & 4.71886(17)  & 4.6483(19)   & 4.6223(15)   & 4.549(2)     & 4.472(2)    \\
c (Å)                         & 7.6412(3)    & 7.486(4)     & 7.431(3)     & 7.283(4)     & 7.135(4)    \\
Volume (Å\textsuperscript{3})              & 147.356(10)  & 140.09(11)   & 137.49(8)    & 130.54(10)   & 123.57(11)  \\
Meas. Reflections; R\textsubscript{int}     & 313; 12.98   & 109; 5.78    & 121; 5.96    & 144; 7.76    & 131; 5.18   \\
No. of indep. reflections     & 28           & 56           & 71           & 119          & 80          \\
R\textsubscript{1}; wR$_2$ (all intensities) & 10.88; 23.57 & 12.29; 33.32 & 17.38; 44.00 & 12.59; 30.12 & 14.99; 48.20\\
GoF                          & 1.4147       & 1.8207       & 1.4868       & 1.5871       & 2.7204      \\ \bottomrule
\end{tabularx}
\end{table}

%%%
\begin{table}[!htbp]
\caption{Crystallographic data for the trigonal YbMg$_2$Bi$_2$ structure at different pressures from single crystal XRD experiments.}
\label{tab:crystallography_YbMg2Bi2} 
\centering
\setlength\tabcolsep{1.5pt}
\begin{tabularx}{\textwidth}{l*{5}{Y}}
\toprule
Pressure (GPa)                & AP         & \round{3.36}        & \round{5.21}        & \round{7.06}        & \round{10.03}      \\ \midrule
Space Group; Z                & $P\mathrm{\bar{3}}m\mathrm{1}$; 1  & $P\mathrm{\bar{3}}m\mathrm{1}$; 1   & $P\mathrm{\bar{3}}m\mathrm{1}$; 1   & $P\mathrm{\bar{3}}m\mathrm{1}$; 1   & $P\mathrm{\bar{3}}m\mathrm{1}$; 1   \\
a (Å)                         & 4.7149(4)  & 4.6502(11)  & 4.6034(4)   & 4.5624(6)   & 4.5058(8)  \\
b (Å)                         & 4.7149(4)  & 4.6502(11)  & 4.6034(4)   & 4.5624(6)   & 4.5058(8)  \\
c (Å)                         & 7.6063(7)  & 7.474(3)    & 7.3782(7)   & 7.2937(10)  & 7.1702(13) \\
Volume (Å\textsuperscript{3})              & 146.43(2)  & 139.98(7)   & 135.406(19) & 131.48(3)   & 126.07(4)  \\
Meas. Reflections; R\textsubscript{int}    & 152; 6.66  & 65; 2.42    & 126; 3.18   & 123; 2.80   & 126; 3.08  \\
No. of indep. reflections     & 1096       & 161         & 347         & 324         & 315        \\
R\textsubscript{1}; wR$_2$ (all intensities) & 2.99; 7.05 & 3.83; 11.58 & 4.94; 27.52 & 5.71; 18.17 & 7.09; 20.31\\
GoF                          & 1.106      & 1.369       & 1.422       & 1.228       & 1.286       \\ \bottomrule
\end{tabularx}
\end{table}

%%%
\begin{table}[!htbp]
\caption{Crystallographic data for the monoclinic phase of CaMg$_2$Bi$_2$ and YbMg$_2$Bi$_2$.}
\label{tab:crystallography_HP} 
\centering
\setlength\tabcolsep{1.5pt}
\begin{tabularx}{\textwidth}{l*{5}{Y}}
\toprule
                              & CaMg$_2$Bi$_2$    & YbMg$_2$Bi$_2$     \\ \midrule
                              
Pressure (GPa)                & \round{12.82}       & \round{11.61}        \\ 
Space Group; Z                & $C\mathrm{2}/m$; 2     & $C\mathrm{2}/m$; 2      \\
a (Å)                         & 7.65(3)     & 7.558(7)     \\
b (Å)                         & 4.333(3)    & 4.489(6)     \\
c (Å)                         & 7.082(10)   & 7.081(6)     \\
$\beta$ ($\mathrm{^\circ}$)                      & 94.36(4)    & 91.82(7)     \\
Volume (Å\textsuperscript{3})              & 234.2(9)    & 240.1(4)     \\
Meas. Reflections; R\textsubscript{int}     & 70; 1.54    & 128; 3.90    \\
No. of indep. Reflections     & 69          & 290          \\
R\textsubscript{1}; wR$_2$ (all intensities) & 9.60; 26.20 & 18.31; 50.16 \\
GoF                          & 1.525       & 2.043       \\ \bottomrule
\end{tabularx}
\end{table}

%%%
\begin{table}[!htbp]
\caption{\textcolor{black}{Structure parameters of the high-pressure structure for CaMg$_2$Bi$_2$ at 12.8 GPa. Note that the uncertainty for the Mg atom is significantly higher than those observed for Ca and Bi. These elevated values strongly suggests significant disorder or positional uncertainty for the Mg atom.}}
\centering
\begin{tabular}{cccccccc}
\toprule
Atom & x & y & z & Occ. &  U$_{eq}$ (\AA$^2$) & Site & Sym. \\
\midrule
Ca &  1.000(3) & 1.000(3) & 0.000(1) & 1.000 & 0.020 & 2a & 2/m \\
Bi &   0.648(2) & 1.000(3) &  0.253(1) & 1.000 & 0.028 & 4i & m \\
Mg &   0.850(17) & 0.500(30) & 0.385(11) & 1.000 & 0.030 & 4i & m \\
\bottomrule
\end{tabular}
\end{table}

%%%
\begin{table}[!htbp]
\caption{\textcolor{black}{Structure parameters of the high-pressure structure for YbMg$_2$Bi$_2$ at 11.6 GPa}}
\centering
\begin{tabular}{cccccccc}
\toprule
Atom & x & y & z & Occ. & U$_{eq}$ (\AA$^2$) & Site & Sym. \\
\midrule
Yb & 0.500(1) & 0.500(1) & 1.000(1) & 1.000 & 0.050 & 2a & 2/m \\
Mg & 0.161(6) & 0.500(1) & 0.647(4) & 1.000 & 0.037 & 4i & m \\
Bi & 0.172(1) & 0.500(1) & 0.239(1) & 1.000 & 0.050 & 4i & m \\
\bottomrule
\end{tabular}
\end{table}

\begin{figure}[!htbp]
    \centering
    \includegraphics[width=0.85\textwidth]{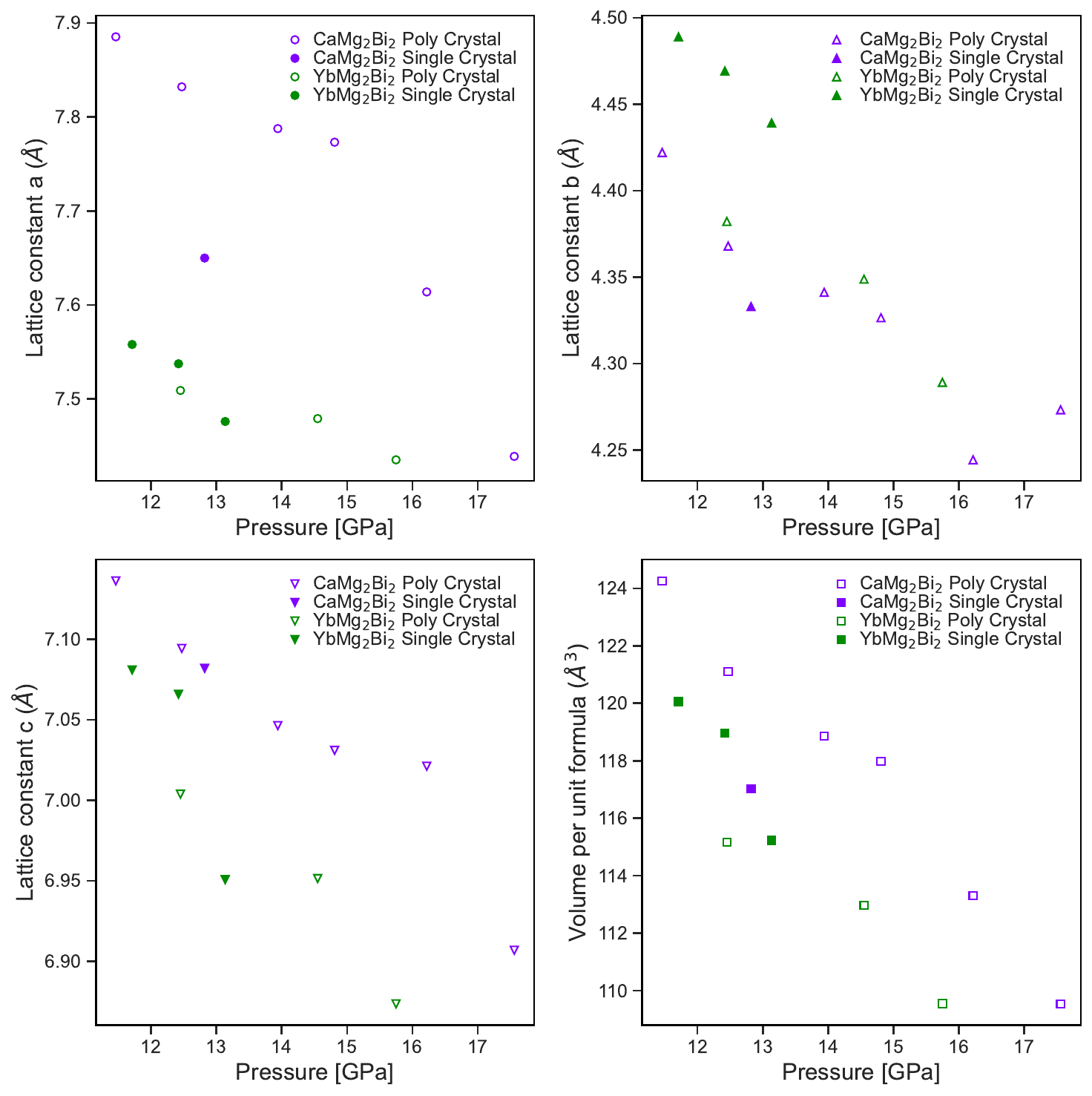}
    \caption{\textcolor{black}{Lattice parameters for the $C2/m$ structure above the phase transition for the single- and poly-crystal available data. Note that the volume per unit formula is shown.}}
    \label{SI_fig:HP_lat_params}
\end{figure}

\subsection{Volume calculation of polyhedra and error estimations}
\label{SI_sec:error}
The volumes of the octahedra and tetrahedra were calculated by treating them as regular polyhedra, using an average of the two distinct Bi-Bi bond distances as the edge length of the octahedron ($\mathrm{l_{oct}}$) and of the tetrahedron ($\mathrm{l_{tet}}$). The octahedron ($\mathrm{V_{oct}}$) and tetrahedron ($\mathrm{V_{tet}}$) volumes are then calculated as:
\begin{equation}
V_{oct}=\frac{\sqrt{2}l_{oct}^3}{3}
    \label{eq:V_Octa}
\end{equation}
and
\begin{equation}
V_{tet}=\frac{l_{tet}^3}{6\sqrt{2}}
    \label{eq:V_Tetra}
\end{equation}
respectively. To calculate the uncertainty in the polyhedral volume, an error propagation scheme was calculated that starts with the uncertainty in the bond lengths as is explained next. For the octahedron volume, the longer Bi-Bi bond length ($\mathrm{l_{{oct}_{I}}}$) and the corresponding uncertainty of such bond ($\mathrm{u_{{oct}_{I}}}$) was used to determine the maximum octahedral volume $\mathrm{V_{{oct}_{max}}}$ as:
\begin{equation}
V_{{oct}_{max}}=\frac{\sqrt{2}\left(l_{{oct}_{I}}+u_{{oct}_{I}}\right)^3}{3}
    \label{eq:V_Octa_max}
\end{equation}
Similarly, the minimum octahedron volume $\mathrm{V_{{oct}_{max}}}$is calculated using the shorter Bi-Bi bond length ($\mathrm{l_{{oct}_{II}}}$) and its associated uncertainty ($\mathrm{u_{{oct}_{II}}}$):
\begin{equation}
    V_{{oct}_{min}}=\frac{\sqrt{2}\left(l_{{oct}_{II}}-u_{{oct}_{II}}\right)^3}{3}
     \label{eq:V_Octa_min}
\end{equation}
Likewise, the maximum tetrahedron volume ($\mathrm{V_{{tet}_{max}}}$) is calculated as a regular polyhedron via the longer Bi-Bi bond ($\mathrm{l_{{tet}_{I}}}$) and its uncertainty ($\mathrm{u_{{tet}_{I}}}$), as a modification of equation \ref{eq:V_Octa}. It becomes:
\begin{equation}
V_{{tet}_{max}}=\frac{\left(l_{{tet}_{I}}-u_{{tet}_{I}}\right)^3}{6\sqrt{2}}
    \label{eq:V_Tetra_max}
\end{equation}
Conversely, the minimum tetrahedron volume ($\mathrm{V_{{tet}_{min}}}$) is based on the smaller Bi-Bi bond that spans the polyhedron ($\mathrm{l_{{tet}_{II}}=l_{{oct}_{I}}}$) and its corresponding uncertainty ($\mathrm{u_{{tet}_{II}}=u_{{oct}_{I}}}$), which is then:
\begin{equation}
\mathrm{V_{{tet}_{min}}=\frac{\left(l_{{tet}_{II}}-u_{{tet}_{II}}\right)^3}{6\sqrt{2}}=\frac{\left(l_{{oct}_{I}}-u_{{oct}_{I}}\right)^3}{6\sqrt{2}}} 
\end{equation}
which, again, is a modification of equation \ref{eq:V_Tetra}. \par
Now, in order to calculate the total polyhedral volume error ($\mathrm{e_{Vol}}$), the \textit{symmetric} mean absolute percentage error is obtained with the following general relation:
\begin{equation}
e_{V}=\frac{1}{n}\sum_{i=1}^n\left|\frac{V_{avg}-V_i}{2\left(V_{avg}+V_i\right)}\right| 
\label{eq:MAPE}
\end{equation}
where $\mathrm{V_{avg}}$ is the average volume and $\mathrm{V_{i}}$the extreme volumes, i.e., either $\mathrm{V_{max}}$ or $\mathrm{V_{min}}$. Then, using equation\ref{eq:MAPE} the octahedral $\mathrm{\left(e_{V_{{oct}}}\right)}$ and tetrahedral volume $\mathrm{\left(e_{V_{{tet}}}\right)}$ errors become:
\begin{equation}
    e_{V_{\text{oct}}}=\frac{1}{2}\left[\left|\frac{V_{\text{oct}}-V_{\text{oct}_{\text{max}}}}{2\left(V_{\text{oct}}+V_{\text{oct}_{\text{max}}}\right)}\right|+\left|\frac{V_{\text{oct}}-V_{\text{oct}_{\text{min}}}}{2\left(V_{\text{oct}}+V_{\text{oct}_{\text{min}}}\right)}\right|\right]
\label{eq:error_V_O}
\end{equation}
and
\begin{equation}
    e_{V_{\text{tet}}}=\frac{1}{2}\left[\left|\frac{V_{\text{tet}}-V_{\text{tet}_{\text{max}}}}{2\left(V_{\text{tet}}+V_{\text{tet}_{\text{max}}}\right)}\right|+\left|\frac{V_{\text{tet}}-V_{\text{tet}_{\text{min}}}}{2\left(V_{\text{tet}}+V_{\text{tet}_{\text{min}}}\right)}\right|\right]
\label{eq:error_V_T}
\end{equation}
respectively.
\clearpage
\subsection{Rietveld refinements}
\begin{figure}[!htbp]
    \centering
    \captionsetup[subfigure]{justification=raggedright,singlelinecheck=false}
    \begin{subfigure}{0.85\textwidth}
        \caption{}
        \centering
        \includegraphics[width=0.85\textwidth]{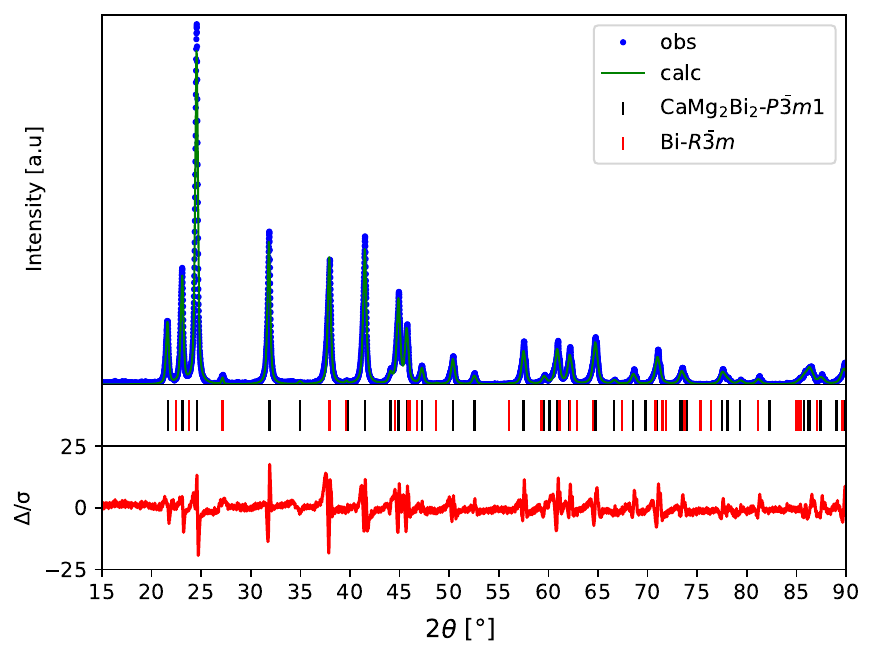}
        \label{SI_fig:Ca_AP_Rietveld}
    \end{subfigure}%
    
    \vspace{-2em} 
    
    \begin{subfigure}{0.85\textwidth}
        \caption{}
        \centering
        \includegraphics[width=0.85\textwidth]{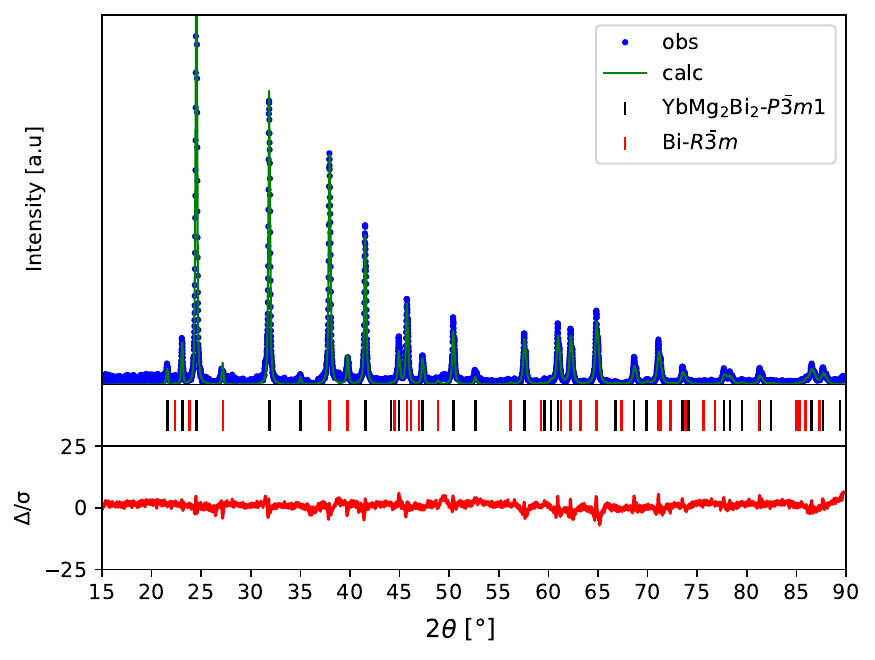}
        \label{SI_fig:Yb_AP_Rietveld}
    \end{subfigure}
    \caption{\textcolor{black}{Ambient-pressure ($P\mathrm{\bar{3}}m\mathrm{1}$) Rietveld refinements of \subref*{SI_fig:Ca_AP_Rietveld}) CaMg$_2$Bi$_2$ and \subref*{SI_fig:Yb_AP_Rietveld}) YbMg$_2$Bi$_2$. In \subref*{SI_fig:Ca_AP_Rietveld}), the Bi impurity phase is around 3\%. In \subref*{SI_fig:Yb_AP_Rietveld}), the content of Bi is approx. 5\%.}}
    \label{SI_fig:Rieveld_AP}
\end{figure}
%%%
\begin{figure}[!htbp]
    \centering
    \captionsetup[subfigure]{justification=raggedright,singlelinecheck=false}
    \begin{subfigure}{\textwidth}
        \caption{}
        \centering
        \includegraphics[width=0.85\textwidth]{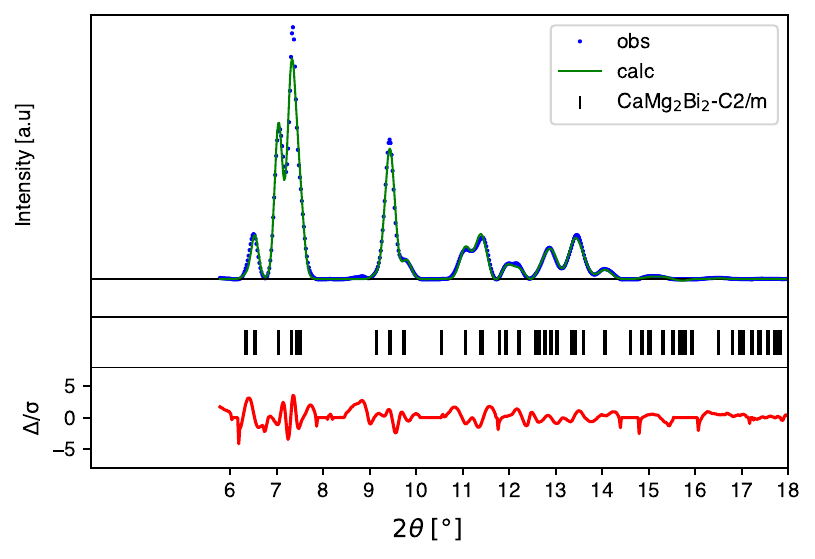}
        \label{SI_fig:Ca_HP_Rietveld}
    \end{subfigure}
    
    \vspace{-2em} 
    
    \begin{subfigure}{\textwidth}
        \caption{}
        \centering
        \includegraphics[width=0.85\textwidth]{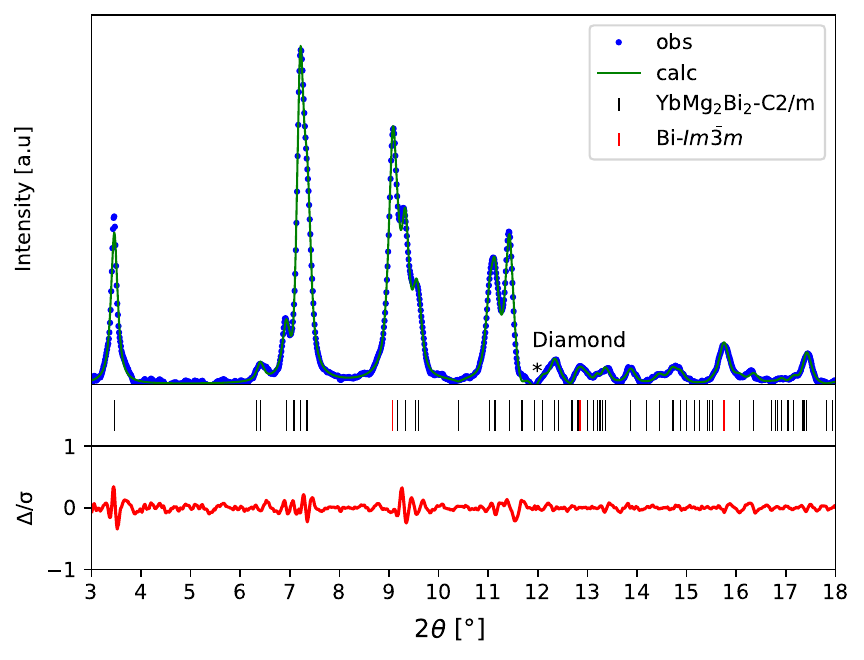}
        \label{SI_fig:Yb_HP_Rietveld}
    \end{subfigure}
    \vspace{-2em} 
    \caption{High-pressure ($C$2/$m$) Rietveld refinements for \subref*{SI_fig:Ca_HP_Rietveld}) CaMg$_2$Bi$_2$ at 12.47 GPa and \subref*{SI_fig:Yb_HP_Rietveld}) YbMg$_2$Bi$_2$ at 12.45 GPa.}
    \label{SI_fig:Rieveld_HP}
\end{figure}
\end{document}